\documentclass{elsart}

 \usepackage{graphics}

\usepackage{amsmath,amssymb,graphicx}

\begin{document}

\begin{frontmatter}



\title{Rescue Model for the Bystanders' Intervention in Emergencies}
\author{Hang-Hyun Jo\corauthref{cor1}}\ead{kyauou2@kaist.ac.kr},
\author{Woo-Sung Jung}, and
\author{Hie-Tae Moon} \corauth[cor1]{Corresponding author. Fax:
+82-42-869-2510.}

\address{Department of Physics, Korea Advanced Institute of Science and Technology,
Daejeon 305-701, Republic of Korea}


\begin{abstract}

To investigate an effect of social interaction on the bystanders'
intervention in emergency situations we introduce a rescue model
which includes the effects of the victim's acquaintance with
bystanders and those among bystanders. This model reproduces the
surprising experimental result that the helping rate tends to
decrease although the number of bystanders $k$ increases. The model
also shows that given the coupling effect among bystanders, for a
certain range of small $k$ the helping rate increases according to
$k$ and that coupling effect plays both positive and negative roles
in emergencies. Finally we find a broad range of coupling strength
to maximize the helping rate.

\end{abstract}

\begin{keyword}
Bystander effect, Social impact theory, Collective phenomena
\PACS 87.23.Ge \sep 89.65.-s \sep 89.90.+n
\end{keyword}
\end{frontmatter}

\section{Introduction}

Recently the concepts and methods of statistical physics and
nonlinear dynamics are applied to investigate social, economic and
psychological phenomena
\cite{Weidlich1991,Mantegna2000,Vallacher1994}. Among the
interesting subjects that have attracted physicists are the opinion
dynamics \cite{Stauffer2005,Sznajd2000} including voting process
\cite{Alves2002,Bernardes2002} and social impact theory
\cite{Latane1981,Nowak1990,Lewenstein1992,Kohring1996,Plewczynski1998,Holyst2000}.
Social impact theory stemmed from the bystander effect that was
first demonstrated in the laboratory by psychologists who studied
the effect of social interaction among bystanders \cite{Latane1969}.
This effect is a marvel phenomenon where people are less likely to
intervene in emergencies when others are present than when they are
alone. A well known example is Kitty Genovese case
\cite{wikipediaKitty}. She was stabbed to death at one night in 1964
by a mentally ill murderer for about 30 minutes while at least 38
witnesses were believed to have failed to help the victim. On the
other hand the refutation that the popular account of the murder is
mostly wrong was suggested on the website \cite{DeMay2005}.
Regardless to this controversy \cite{Rasenberger2004} many
experimental results in the laboratory and from field studies show
the bystander effect obviously \cite{Latane1969,Amato1983}.

Although the social impact theory originated from the bystander
effect and related models have been developed, there does not exist
any model describing quantitatively the bystander effect itself. To
construct the model mentioned above we abstract a few important
factors from the existing literatures
\cite{Latane1969,Piliavin1982}. The main dependent variables are the
rates of reporting the smoke in the test room, helping endangered
victims and so on. When tested alone, subjects behaved reasonably
and the response rate was high. However the rate was significantly
depressed when they were with other subjects. Subjects with others
were unsure of what had happened or thought other people would or
could do something. In another experiment subjects who were friends
responded faster than those who were strangers. The subjects who had
met the victim were significantly faster to report victim's distress
than other subjects. And the degree of arousal that bystanders
perceive is a monotonic positive function of the perceived severity
and clarity of the emergency, and bystander's emotional involvement
with the victim, which will be also considered by our rescue model
in an abstract way.

\section{Rescue Model}

A simple rescue model is introduced for investigating the
statistical feature of the effects of the victim's acquaintance with
bystanders and those among bystanders in emergency situations. We
focus on the relations between agents rather than agents themselves,
so define a relation spin between two agents as $a_{ij}$ which is
$1$ if agents $i$ and $j$ have been successful in the intervention
and $0$ otherwise. $a_{ij}$ can be interpreted as an element of
adjacency matrix of helping network. Each agent $i$ has its
intervention threshold $c_i$ over which that agent can try to
intervene in an emergency situation.

At each time step an accident happens which consists of the degree
of accident represented as a random number $q_v$ uniformly drawn
from $[0,1]$, a randomly selected victim $v$, the number of
bystanders $k$ confined to $[1,N-1]$, and a set of randomly selected
$k$ bystanders $N_v$ from a population. Then the update rule is as
following:
\begin{equation}
a_{vi}(t+1)=\theta \left(q_v+\alpha a_{vi}(t) +\beta \sum_{j\in N_v,
j\neq i}{(2a_{ij}(t)-1)} -c_i\right)
\end{equation}
where $\theta(x_i)$ is a heaviside step function. Only one bystander
$i\in N_v$ with the largest value $x_i$ can intervene per accident.
If we assume that the response speed of bystander $i$ is exponential
in $x_i$, the selection of the bystander with the largest $x_i$ is
justified. Additionally, once one bystander intervenes, the
pressures on the others will disappear. If $x_i\geq0$, the rescue
succeeds and then the victim $v$ and that bystander $i$ gain
acquaintances if they have not been related to each other. In case
of $x_i<0$ the rescue fails and then their acquaintance is cut if
existed. $\alpha$ represents the degree of victim's acquaintance
with bystander, so can be called an acquaintance strength. The third
term of $\theta$ function is related to the acquaintances among
bystanders, for each relation $2a_{ij}-1$ gives $1$ if two
bystanders know each other or $-1$ otherwise. There does not exist
any neutral relation here. $\beta$ is used to tune the strength of
coupling so can be called a coupling strength. As an order parameter
we adopt the average helping rate:
\begin{equation}
\langle a\rangle(t)=\frac{2}{N(N-1)}\sum_{i<j}{a_{ij}(t)}
\end{equation}
which can be also interpreted as a social temperature or an average
linkage of the helping network. We fix $c_i\equiv c=0.25$ for all
$i$ according to the experimental result \cite{Latane1969} that
$70\sim 75\%$ of alone subjects intervened and $c$ does not change
through this paper, which means we consider a population composed of
homogeneous and non-adaptive agents. And for the most cases $\alpha$
is also fixed as $0.1$. The main control parameter is the number of
bystanders $k$. Finally, the initial conditions are $a_{ij}=0$ for
all pairs.

\subsection{Case with $\beta=0$}

At first let us consider only the effect of victim's acquaintance
with bystanders and ignore those among bystanders. Generally an
equation for the average helping rate can be written as
\begin{equation}
\frac{da_k(t)}{dt}=W_{0\rightarrow1}-W_{1\rightarrow0}
\end{equation}
where $a_k(t)$ denote $\langle a\rangle(t)$ for a fixed $k$. Given
the values of parameters as above, when $0\leq q_v <c-\alpha$, the
rescue fails independent of $a_{vi}(t)$ so one of acquaintances, if
exists, should be cut. When $c-\alpha \leq q_v <c$,
$a_{vi}(t+1)=a_{vi}(t)$ so this interval of $q_v$ does not
contribute to the equation. When $c\leq q_v \leq 1$ the rescue
succeeds independent of $a_{vi}(t)$ so a new acquaintance is formed
only if there is no acquaintance between them.
\begin{eqnarray}
W_{0\rightarrow1}&=&(1-c)(1-a_k(t))^k \\
W_{1\rightarrow0}&=&(c-\alpha)\left(1-(1-a_k(t))^k\right)
\end{eqnarray}
The stationary solution is obtained easily.
\begin{equation}
a_k=1-\left(\frac{c-\alpha}{1-\alpha}\right)^{1/k}
\end{equation}
The solution means that although the coupling strength is not
considered ($\beta=0$), the helping rate depends on the number of
bystanders $k$. As increasing $k$ the probability that the victim
does not know any bystanders (contributes to $W_{0\rightarrow 1}$)
decreases rapidly and the complementary probability (contributes to
$W_{1\rightarrow 0}$) increases. This is why the helping rate
decays. In this case $\alpha$ affects not the decaying behavior but
the decaying speed only when $\alpha<c$. If $\alpha$ becomes larger
than $c$, $W_{1\rightarrow 0}=0$ so $a_k=1$ independent of $k$.

\begin{figure}
\includegraphics[scale=.9]{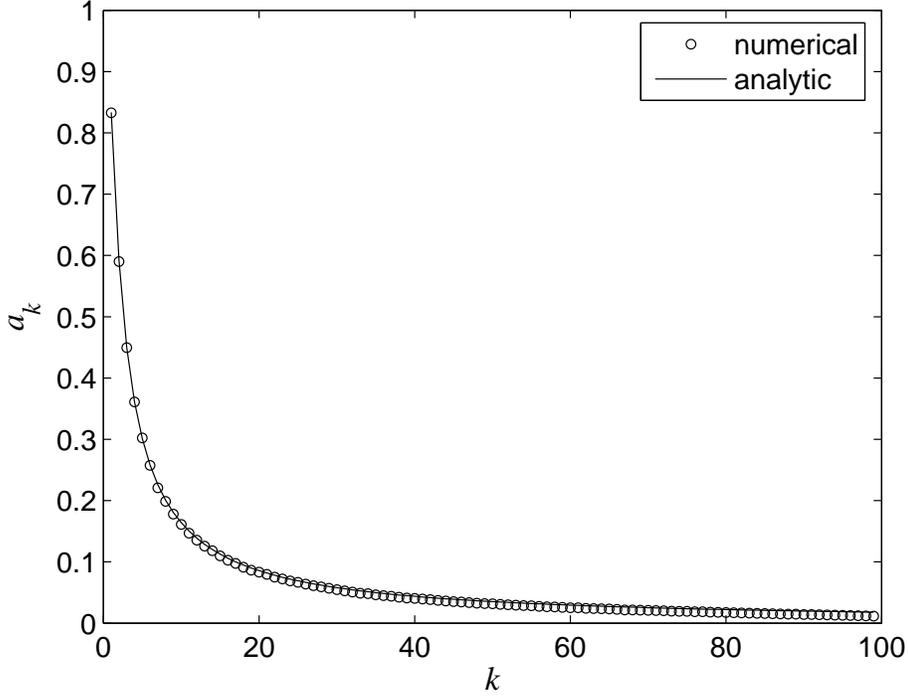}
\caption{The analytic solution and numerical simulations of the case
with $\alpha>0$ and $\beta=0$. Here $N=100$, $c=0.25$ and
$\alpha=0.1$.} \label{fig:1}
\end{figure}

The numerical simulations are shown in Fig. \ref{fig:1} and
reproduce the experimental results that people who live in larger
cities are less likely to help the strangers \cite{Amato1983}.
Compared to Ref. \cite{Amato1983} $k$ corresponds to the size of
city or town.

\subsection{Case with $\beta>0$ and $k=2$}

If the coupling among bystanders is taken into account and $k=2$,
the interval $[0,1]$ of $q_v$ is divided into five subintervals by
four points $c-\alpha-\beta$, $c-\alpha+\beta$, $c-\beta$, and
$c+\beta$. Whether $c-\alpha+\beta$ is smaller or larger than
$c-\beta$ does not affect the result. For each subinterval the
transition rate is calculated.
\begin{eqnarray}
W_{0\rightarrow1}&=&(1-c-\beta)(1-a_2)^3+(1-c+\beta)a_2(1-a_2)^2 \\
W_{1\rightarrow0}&=&(c-\alpha+\beta)a_2(1-a_2)(2-a_2)+(c-\alpha-\beta)a_2^2(2-a_2)
\end{eqnarray}
The second term in R.H.S. of $W_{1\rightarrow0}$ vanishes when
$c-\alpha-\beta<0$, so we get the solutions for both cases.
\begin{equation}
a_2=\left\{ \begin{array}{ll}
\frac{1-\alpha-\beta-\sqrt{\alpha^2+\beta^2+\alpha\beta-\alpha-\beta+c(1-\alpha)}}
{1-\alpha} & \textrm{if $c-\alpha-\beta\geq0$}\\
\frac{1+c-2\alpha-\beta-\sqrt{5c^2+4\alpha^2-3\beta^2-8\alpha
c-2\beta c-2c+2\beta+1}}{2(c-\alpha-\beta)} & \textrm{if
$c-\alpha-\beta<0$}
\end{array}\right.
\end{equation}

\begin{figure}
\includegraphics[scale=.9]{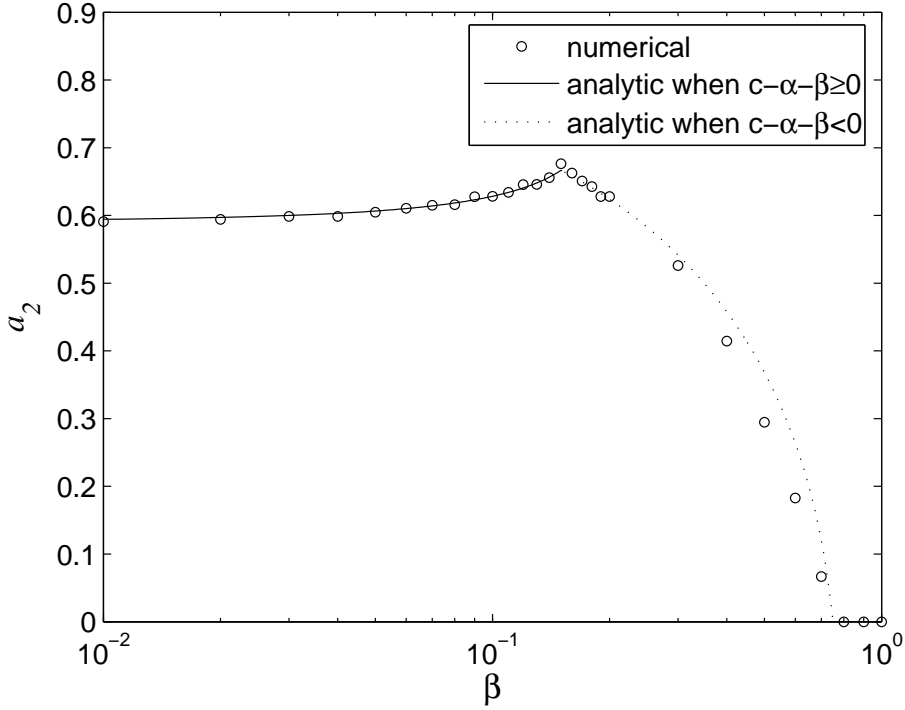}
\caption{The analytic solution and numerical simulations of the case
with $\alpha,\beta>0$ and $k=2$. $a_2$ has a maximum value when
$c-\alpha-\beta=0$. The other parameter values are the same as Fig.
\ref{fig:1}.} \label{fig:2}
\end{figure}

The numerical simulations in Fig. \ref{fig:2} support the analytic
solution. One can find a maximum value of $a_2$ when
$c-\alpha-\beta=0$, which means that there exists an optimal
coupling strength $\beta_{opt}=c-\alpha$ to maximize the helping
rate when given $c$ and $\alpha$.

\subsection{General case with $\beta>0$}

For a general $k$, the interval $[0,1]$ of $q_v$ is divided by the
following points: $c-\alpha-(k-1)\beta$, $\cdots$,
$c-\alpha+(k-1)\beta$, $c-(k-1)\beta$, $\cdots$, $c+(k-1)\beta$. For
each subinterval the transition rate can be calculated.
\begin{eqnarray}
W_{0\rightarrow1}&=&(1-a_k)^k\left(1-c+(k-1)\beta-2\beta F(a_k,k)\right) \\
W_{1\rightarrow0}&=&\left(1-(1-a_k)^k\right)\left(c-\alpha-(k-1)\beta+2\beta
F(a_k,k)\right)
\end{eqnarray}
where
\begin{equation}
F(a_k,k)=\sum_{n=0}^{k-2}(k-1-n)\left(\begin{array}{c}\frac{1}{2}k(k-1)
\\n\end{array}\right) a_k^n(1-a_k)^{\frac{1}{2}k(k-1)-n}
\end{equation}
Consequently to get a stationary solution we must solve the
following equation.
\begin{equation}
(1-\alpha)(1-a_k)^k=c-\alpha-(k-1)\beta+2\beta F(a_k,k)
\label{eq:com}
\end{equation}
Since $F(a_k,k)$ is the $\frac{1}{2}k(k-1)$th order polynomial, the
order of above equation is $\max(\frac{1}{2}k(k-1),k)$ and difficult
to solve exactly for all $k$. Therefore, as an approximation we
consider only the most left, the second and the most right
subintervals.
\begin{eqnarray}
W_{0\rightarrow1}&=&\left(1-c-(k-1)\beta\right)(1-a_k)^k \\
W_{1\rightarrow0}&=&\left(c-\alpha-(k-3)\beta\right)\left(1-(1-a_k)^k\right)
\end{eqnarray}
The approximate stationary solution is
\begin{equation}
a_k=1-\left(\frac{c-\alpha-(k-3)\beta}{1-\alpha-2(k-2)\beta}\right)^{1/k}
\textrm{when $k\leq k_1\equiv \frac{c-\alpha}{\beta}+3$.}
\label{eq:11}
\end{equation}

\begin{figure}
\includegraphics[scale=.9]{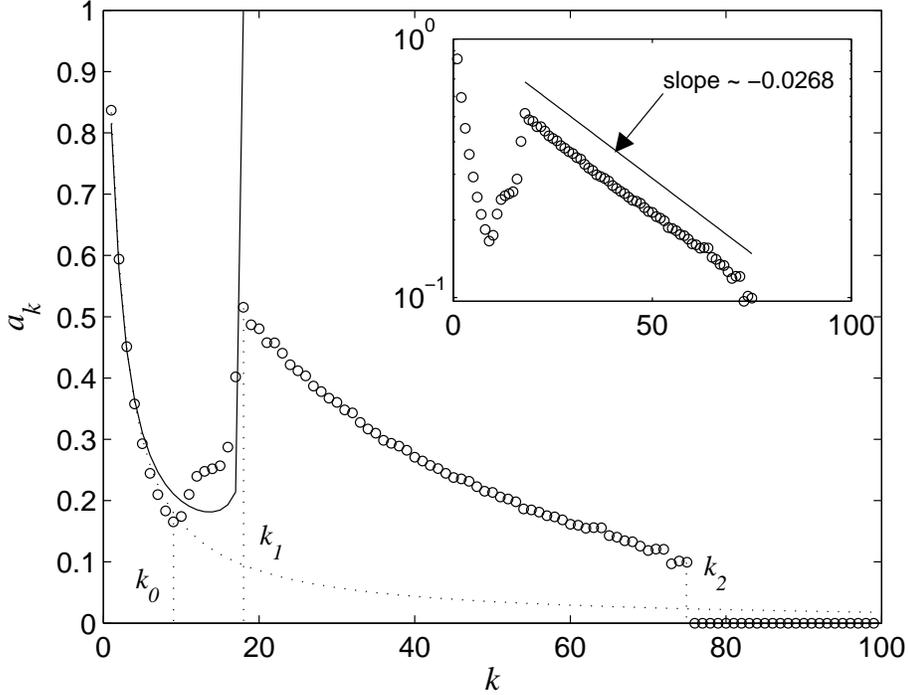}
\caption{The analytic solution and numerical simulations of the case
with $\beta=0.01$. Circles represent the numerical results and solid
line and dotted line do the approximate solution for $\beta=0.01$
and for $\beta=0$ for comparison respectively. Inset graph shows the
exponentially decaying behavior of the same data.} \label{fig:3}
\end{figure}

Figure \ref{fig:3} shows the numerical simulations supporting the
approximation except for $k\approx k_1$. The helping rate undergoes
a trough for $k$ less than $k_1$, which is also partly observed in
the experiment results \cite{Amato1983}, and then decreases
monotonically as for the case without a coupling effect ($\beta=0$).
For $k\geq k_2$ the helping rate shows a sudden drop to $0$. To
understand this behavior intuitively two factors should be
considered. On the one hand as for the case without a coupling
effect, the increasing number of bystanders increases the
probability of cutting acquaintance too. On the other hand the most
left subinterval $[0,c-\alpha-(k-1)\beta)$, which most contributes
to $W_{1\rightarrow 0}$, decreases as increasing $k$. In this paper
$c$ is fixed as a rather small value so that a change of the most
right subinterval rarely affects $W_{0\rightarrow 1}$. In
conclusion, when $k<k_0$ the former factor dominates the latter so
that the helping rate decreases. When $k_0\leq k<k_1$ the increasing
$k$ minimizes the inhibiting effect by other bystanders to make the
helping rate increase. Once $k$ passes $k_1$, the decreasing
subinterval vanishes and only the former factor works.

\begin{figure}
\includegraphics[scale=.9]{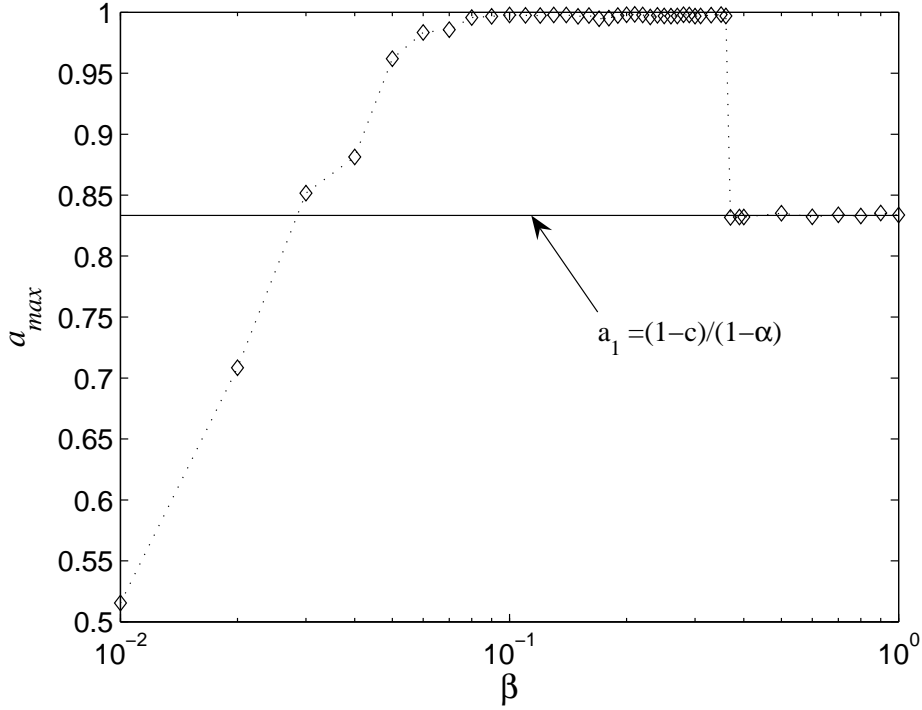}
\caption{The maximum value $a_{max}$ as a function of $\beta$. There
exists a broad range of $\beta$ to maximize the helping rate.}
\label{fig:4}
\end{figure}

Coupling effect among bystanders plays both positive and negative
roles in emergencies. When $k<k_0$, the helping rate is rarely
affected by the coupling. In case of $k_0\leq k\leq k_2$, the
coupling helps the helping rate obviously. Especially when $k_1 \leq
k \leq k_2$, $a_k$ decays exponentially with slope of $-0.0268$ in
the semi-log plot. Once $k$ passes $k_2$, $a_k$ turns to $0$ because
of too many other inhibiting bystanders. The value of $k_2$ can be
obtained by a simple mean field approximation and by using the
following:
\begin{equation}
\langle \theta(x-x_0)\rangle_x=\left\{ \begin{array}{ll}
1 & \textrm{if $x_0\leq0$}\\
1-x_0 & \textrm{if $0< x_0 \leq1$}\\
0 & \textrm{if $x_0 > 1$}
\end{array}\right.
\end{equation}
$\langle\cdot\rangle_x$ means an average over $x\in [0,1]$. From
Equation (1), if $c-\alpha a_k -\beta(k-1)(2a_k-1)>1$, then $a_k=0$,
therefore the condition for $a_k=0$ is $k>k_2\equiv
\frac{1-c}{\beta}+1$.

Under what conditions is the helping rate maximized? For each
$(\alpha, \beta)$ we can find the maximum value of $a_k$, denoted by
$a_{max}$. It is sufficient to compare $a_1$ to $a_{k_1}$, where
$k_1$ is defined by Eq. \ref{eq:11}. $a_1$ is $\frac{1-c}{1-\alpha}$
independent of $\beta$ while $a_{k_1}$ is smaller than $a_1$ for
small $\beta$ and exceeds $a_1$ for some range of $\beta$ and then
vanishes, in fact $k_1=1$ for larger $\beta$ in Fig. \ref{fig:4}.
Especially $a_{max}$ approaches to $1$ when $0.08\leq\beta\leq0.35$.
There exists a broad range of $\beta$ for which the helping network
is almost fully connected.


\section{Conclusions}

We introduced a simple rescue model to investigate the effects of
victim's acquaintance with bystanders (acquaintance strength
$\alpha$) and those among bystanders (coupling strength $\beta$) for
the bystanders' intervention in emergency situations. When
$\beta=0$, as the increasing number of bystanders $k$ the helping
rate decreases, where the speed depends on $\alpha$. For the case of
$\beta>0$ and $k=2$, there exists an optimal coupling strength
$\beta=c-\alpha$ for the maximum helping rate. Coupling strength
plays both positive and negative roles in emergencies. For $k_0\leq
k<k_1$, since the coupling among bystanders minimizes the inhibiting
effect, the helping rate increases according to $k$. And then the
helping rate decays monotonically as for the case without a coupling
effect. Once $k$ passes $k_2$ too many bystanders inhibit the
helping. There exists a rather broad range of $\beta$ where almost
all the trials to intervene in emergencies are successful. In this
case it is not necessary to fine-tune the coupling strength to get
the highest helping rate.

\label{}



\end{document}